\documentclass[conference]{IEEEtran}
\ifCLASSINFOpdf
\else
\fi
\usepackage{amsmath}
\usepackage{xcolor}
\usepackage{bbold}
\usepackage{graphicx}
\usepackage{mathtools}
\usepackage{siunitx}
\usepackage{soul}
\DeclarePairedDelimiter{\ceil}{\lceil}{\rceil}

\DeclareMathOperator*{\argmin}{arg\,min}

\begin{document}
%
\title{On the Computational Viability of Quantum Optimization for PMU Placement}


\author{\IEEEauthorblockN{Eric B. Jones}
\IEEEauthorblockA{Department of Physics\\
Colorado School of Mines and\\
National Renewable Energy Laboratory\\
Golden, Colorado 80401, USA\\
Email: Eric.Jones@nrel.gov}
\and
\IEEEauthorblockN{Eliot Kapit}
\IEEEauthorblockA{Department of Physics\\Colorado School of Mines\\
Golden, Colorado 80401, USA\\
Email: Ekapit@mines.edu}
\and
\IEEEauthorblockN{Chin-Yao Chang, David Biagioni,\\ Deepthi Vaidhynathan, Peter Graf,\\ and Wesley Jones}
\IEEEauthorblockA{National Renewable Energy Laboratory\\
Golden, Colorado 80401, USA\\
Email: ChinYao.Chang@nrel.gov}}


%


\maketitle

\begin{abstract}
Using optimal phasor measurement unit placement as a prototypical problem, we assess the computational viability of the current generation D-Wave Systems 2000Q quantum annealer for power systems design problems. We reformulate minimum dominating set for the annealer hardware, solve the reformulation for a standard set of IEEE test systems, and benchmark solution quality and time to solution against the CPLEX Optimizer and simulated annealing. For some problem instances the 2000Q outpaces CPLEX. For instances where the 2000Q underperforms with respect to CPLEX and simulated annealing, we suggest hardware improvements for the next generation of quantum annealers.
\end{abstract}


%
\IEEEpeerreviewmaketitle

\section{Introduction \label{intro}}

As progress continues to be made towards universal error-corrected quantum computers, many opportunities exist to test the problem solving capabilities of current and near-term quantum hardware \cite{preskill2018quantum}. Along with problems in chemistry, artificial intelligence, and sampling, combinatorial optimization problems are excellent candidates to see quantum-enhanced speedups to solution \cite{neill2018blueprint}. 

Meanwhile, a new paradigm is emerging regarding how to construct next-generation energy grids that are secure, resilient, cost-effective, and which can incorporate large quantities of distributed renewable energy. Such systems will likely involve intensive online computation, optimal control over multiple timescales, and extensive state monitoring in order to dynamically adapt to varying generation and demand \cite{kroposki2017autonomous}. Given the complexity of this task, offline optimization and rational design of grid properties that allow more efficient online computation and observation is crucial to the performance of future power networks. In its simplest depiction, a power grid may be modeled as an undirected graph where buses in the system are assigned to graph nodes and branches are assigned to graph edges. At this level of abstraction, the first step towards designing a power system consists of solving a combinatorial optimization problem defined over the graph. Many power grid-relevant combinatorial optimization problems are NP-complete \cite{verma2010power} \cite{schumacher2014optimization} \cite{yang2015risk}. It is therefore important to identify and assess the performance of novel approximate and heuristic solution methods for combinatorial optimization problems particular to power systems design in instances where exact solution is infeasible.

Adiabatic quantum annealing (AQA) constitutes one of the main efforts to outperform classical solution methods on hard combinatorial optimization problems \cite{kadowaki1998quantum}. Definite runtime speedups have been demonstrated for AQA against both simulated annealing (SA)-- in the form of a scaling advantage-- and quantum Monte Carlo (QMC)-- as a fixed prefactor-- using the D-Wave 2X and 2000Q machines on proof-of-principle problems designed to have tall and narrow energy barriers \cite{denchev2016computational} \cite{albash2018demonstration}. Additionally, the largest D-Wave annealer has 2,048 qubits \cite{preskill2018quantum}. Given the large scale of available quantum annealers and their positive prospects for runtime speedup on solving combinatorial optimization problems, we therefore assess the feasibility of speeding up the optimization of power systems using AQA. As a prototypical example, we consider the optimal phasor measurement unit placement (OPMUP) problem. Formulated as a graph theoretic problem, we treat the simplest variant of OPMUP, minimum dominating set (MDS), rather than a more realistic formulation such as power dominating set for clarity in reformulation and discussion. We reformulate the corresponding integer linear program (ILP) into a quantum Hamiltonian operator suitable for solution on a D-Wave quantum annealer. We analyze the scaling of the physical resources required to contend with the MDS formulation on standard Institute of Electrical and Electronics Engineers (IEEE) test power systems ranging in size from 9 to 300 buses. For those problem instances that are minor-embeddable on a (16, 4)-Chimera hardware graph, we assess the ability of the D-Wave 2000Q quantum processing unit (QPU) to accurately find ground state solutions to OPMUP using standard annealing schedules.

The OPMUP problem is applicable to next-generation power system design due to the fact that increased incorporation of renewables into the grid and demand-side management require accurate spatiotemporal state estimation of the grid on sub-second timescales \cite{kroposki2017autonomous}. Synchrophasors, or phasor measurement units (PMUs), are able to measure voltage and current amplitude and phase angles at a rate of 30-60 Hz in a GPS synchronized manner and are therefore able to reconstruct the entire state of the grid if a measurement can be obtained for every bus in the power system \cite{wache2011application}. However, placement of a PMU at every bus is not necessary since PMUs are also able to deduce synchrophasor quantities at adjacent, and in some instances further, buses \cite{brueni2005pmu}. Along with cost considerations, this fact implies that there is a minimum number of PMUs that need to be placed on a given grid topology for full observability, which is substantially fewer than the number of buses in the power system \cite{liao2015hybrid}. The MDS formulation of OPMUP, while being the simplest variant, is nonetheless NP-complete, and there exists a large body of optimization literature devoted to finding its solution \cite{manousakis2011optimal}. And while also residing on the ``highly idealized'' end of the spectrum of design problems future power systems will likely need to surmount, considering the solution of such problems by AQA should point both towards other areas in power system design where quantum optimization could be helpful, and ways in which the attendant quantum hardware could be improved to better address such problems.

\section{Quantum optimization \label{quantop}}

For a discrete set of possible solutions expressed as binary variable strings $\{ x = (x_1 \ldots x_N) \}$, a classical cost function $C(x)$ to be extremized, and a set of constraints $\{ g(x) = 0 \}$, a necessary condition for the combinatorial optimization problem to be adapted to quantum hardware is that the cost function and all constraints in the problem be represented by a quantum Hamiltonian operator, $\hat{H}$ \cite{hadfield2019quantum}. A common procedure for constructing $\hat{H}$ begins by re-expressing the set of constraints as a (usually quadratic) penalty function $P$, which is then added to $C$ in order to write the whole problem as an extremization problem: $H(x) \equiv C(x) + P(x)$. Binary variables $x_i \in \{ 0, 1 \}$ are related to classical spin variables $s_i$ used in SA by the transformation $s_i = 1 - 2 x_i$. A suitable Hamiltonian operator is then obtained by elevating the classical spin variables to single qubit operators, which measure qubit states in the computational basis: $H(s) \rightarrow \hat{H}(\hat{Z})$. Generally, a Hamiltonian obtained this way may be expanded in powers of single qubit operators \cite{denchev2016computational}

\begin{equation} \label{eq:gen_ham}
\hat{H} = - \sum_{k=1}^{K} \sum_{j_1 \ldots j_k = 1}^N J_{j_1 \ldots j_k} \hat{Z}_{j_1} \ldots \hat{Z}_{j_k}.
\end{equation}
In principle, AQA is able heuristically to solve Eq. \eqref{eq:gen_ham} to arbitrary order $K$. However, a particularly relevant class of problem Hamiltonians occurs when $K=2$. Such Hamiltonians fall into the class of problems termed ``quantum Ising models'' and have the form

\begin{equation} \label{eq:is_ham}
\hat{H}_{IS} = - \sum_{j_1=1}^N J_{j_1} \hat{Z}_{j_1} - \sum_{j_1=1}^N \sum_{j_2=1}^N J_{j_1 j_2} \hat{Z}_{j_1} \hat{Z}_{j_2}.
\end{equation}
$\hat{H}_{IS}$ is the quantum analog of the well-known classical Ising model and the equivalent quadratic unconstrained binary optimization (QUBO) problem, and the reason for its prominence is that two-body qubit interactions have been the most straightforward to engineer in hardware with higher-order interactions requiring additional overhead \cite{chancellor2017circuit}. Both the classical Ising model and QUBO fall under the umbrella of binary quadratic models (BQM). Therefore, finding the quantum mechanical ground state to $\hat{H}_{IS}$ is equivalent to solving any combinatorial optimization problem expressed as a BQM. It will be shown in Sec. \ref{sec:opmup} that OPMUP can be formulated as such a BQM and so is amenable to solution on existing quantum hardware. The current generation D-Wave quantum annealer obtains heuristic solutions to Eq. \eqref{eq:is_ham} by evolving the time-dependent Hamiltonian

\begin{equation} \label{eq:dwave_ham}
\hat{H}(t) = - A(t) \sum_{i=1}^N \hat{X}_i + B(t) \hat{H}_{IS}
\end{equation}
adiabatically, through the parameters $A(t)$ and $B(t)$, such that at time $t=0$, $A(0) >> B(0)$ and at time $t=\tau$, $A(\tau) << B(\tau)$, where $\tau$ is the terminal time point in the annealing schedule \cite{denchev2016computational}. Adiabaticity dictates that if evolution according to a time-dependent Hamiltonian is performed slowly enough, then the system upon which the Hamiltonian operates will at every point in time remain in the instantaneous ground state of the Hamiltonian if it was prepared in the ground state of $\hat{H}(0)$ at $t=0$ \cite{kadowaki1998quantum}. The single qubit ground state of each $- \hat{X}_i$ operator is $|+\rangle = (|0\rangle + |1\rangle)/\sqrt{2}$. Therefore, at $t=0$ the annealer assumes the full superposition state

\begin{align}
|\psi(0)\rangle &= |+\rangle^{\otimes N} \nonumber \\
&= \frac{1}{\sqrt{2^N}} \big(|0 \ldots 0\rangle + |0 \ldots 1\rangle + \ldots + |1\ldots1\rangle  \big).
\end{align}
In other words, the initial state of the annealer is an even representation of all solutions to the optimization problem in parallel. As the annealing schedule is carried out, the probability amplitudes that multiply each solution are modulated in order to reflect the superposition, which is the instantaneous ground state so that by the end of the schedule, the resulting state is

\begin{equation}
|\psi(\tau)\rangle = a_1 |0 \ldots 0\rangle + a_2 |0 \ldots 1\rangle + \ldots + a_{2^N} |1\ldots1\rangle,
\end{equation}
where some particular $|a^*|^2$ (or some degenerate set $\{ |a^*|^2 \}$) is (are) now much larger than the rest. In the ideal limit of infinitely long anneal times, $|a^*|^2$ (or $\sum |a^*|^2$) $\rightarrow 1$ \cite{albash2018adiabatic}. Since the probability amplitude that multiplies the optimal solution(s) at the end of the annealing schedule is (are) so much larger than the rest, when the state $|\psi(\tau)\rangle$ is measured, the string that corresponds to the ground state of $\hat{H}_{IS}$ is obtained with high probability.

\section{Reformulation of minimum dominating set for optimal phasor measurement unit placement \label{sec:opmup}}

\begin{figure}[!t]
\centering
\includegraphics[width=0.5\linewidth]{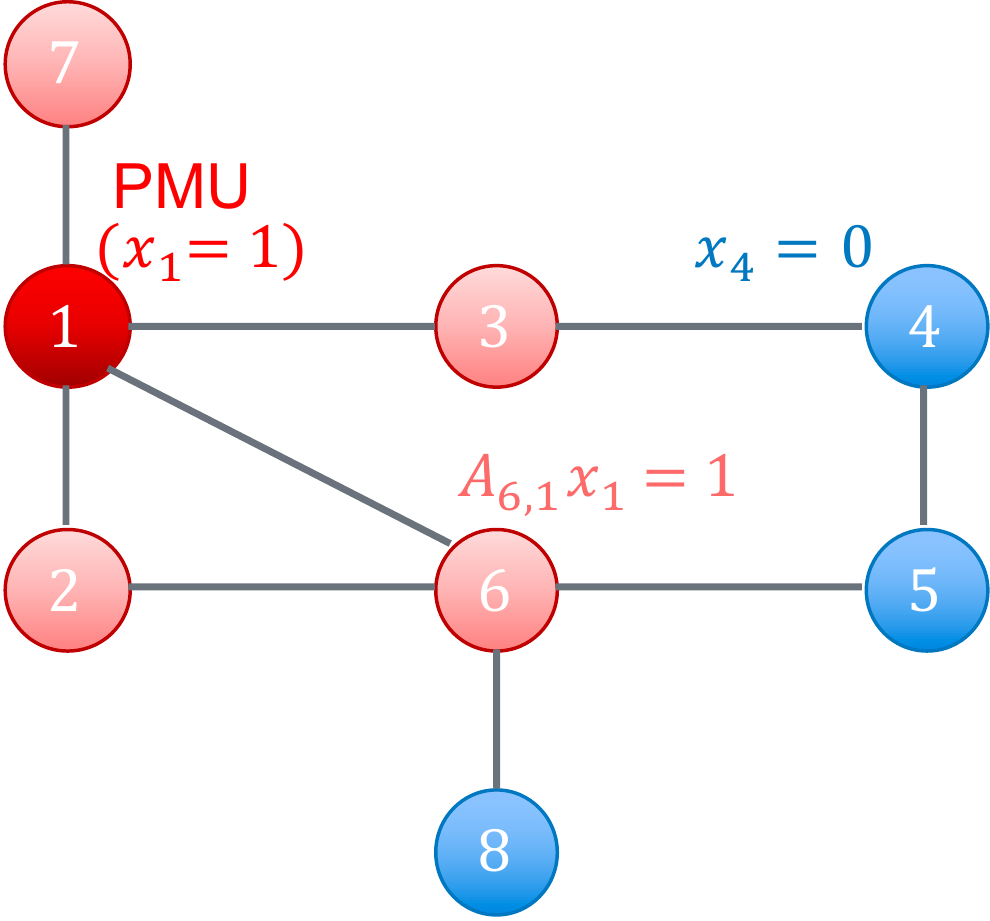}
\caption{MDS formulation of OPMUP. Red is a placed PMU, red and pink are observed nodes as a result, and blue nodes are unobserved. \label{mds_cartoon}}
\end{figure}

The minimum dominating set (MDS) representation of OPMUP can be stated as follows. Let an electric power grid be represented by a graph $G=(V,E)$ where the node set $V$ represents the buses in the system and the edge set $E$ represents the branches. Oftentimes, an edge represents a transmission line, but this is not always the case. We would like to select the minimal initial number of nodes in $V$ (or place the minimal number of PMUs) such that after following the observability rules for the graph, the full graph is observed. Denote the initial selection of nodes as a bit string $x = (x_1 \ldots x_n) \in \{0,1\}^N$ where $0$ denotes an unobserved initial node (no PMU) and $1$ denotes an observed initial node (with PMU). $N=|V|$ is the order of the graph. The simplest set of observability rules we can take are as follows: i) a node is observed if it has had a PMU placed on it and ii) a node is observed if it shares an edge with a PMU. These rules can be summarized using the matrix $A = \tilde{A} + \mathbb{1}$, where $\tilde{A}$ is the adjacency matrix of the graph $G$ and $\mathbb{1}$ is the $N \times N$ identity matrix. Then, if we define column vectors corresponding to the bit strings, $\textbf{x}=(x_1, \ldots, x_n)^T$, and denote $\textbf{b}=(1, \ldots, 1)^{T}$, the appropriate constrained optimization problem is

\begin{equation} \label{cost}
\min \sum_{i=1}^N x_i
\end{equation}

subject to

\begin{equation} \label{constraint}
A\textbf{x} \geq \textbf{b}.
\end{equation}
In Fig. \ref{mds_cartoon}, a PMU has been placed on node 1, indicated in red. As a result, itself along with nodes 2, 3, 6, and 7 are observed (pink). Nodes 4, 5, and 8, in blue, remain unobserved. The minimum number of placed PMUs needed to observe the whole graph is called the domination number, $\gamma(G)$.

Eq. \eqref{cost} indicates that the appropriate cost function to minimize is $C(x)=\sum_i x_i$. Eq. \eqref{constraint} however is not presently in a form that can be enforced as a minimization problem. In order for this to be the case we re-write Eq. \eqref{constraint}  as a quadratic penalty function with $N$ non-negative integer-valued surplus variables $\{ y_i \}$

\begin{equation} \label{penalty_ham}
P(x,y) = \sum_{i=1}^N \alpha_i \Bigg( \sum_{j=1}^N A_{ij} x_j - b_i - y_i \Bigg)^2.
\end{equation}
We note briefly here that while derived independently, our reformulation of minimum dominating set is similar to the treatment by Dinnean and Hua \cite{dinneen2017formulating}. In order to see that $P$ appropriately accounts for Eq. \eqref{constraint}, consider the $i^{th}$ term in $P$. If $\sum_{j=1}^N A_{ij} x_j < b_i$, the minimal value for the term is achieved when $y_i=0$ and a penalty will still be incurred because the square of the term will still be positive. If however, $\sum_{j=1}^N A_{ij} x_j \geq b_i$, $y_i$ can be chosen in order to make the term zero, thus incurring no penalty. Each $y_i$ then needs to be represented as an expansion of binary variables so that it too can be represented on the annealer. To understand the resources required for this, consider that the largest number $y_i$ will ever need to be is $\sum_{j=1}^N A_{ij} (1) - b_i = d_i + 1 - b_i$. Thus, $d_i + 1 - b_i \geq y_i \geq 0$. Let $\mu_i^m \equiv \ceil*{\log_2(d_i+1-b_i)}$, where $\ceil*{}$ is the ceiling function, be the number of bits needed to represent the upper bound to $y_i$ in a binary expansion. Then, the number of ancilla bits needed for the whole problem is $\sum_{i=1}^N \mu_i^m$, the binary expansion of $y_i$ is

\begin{equation}
y_i = \sum_{\mu=0}^{\mu_i^m - 1} 2^{\mu} y_{i \mu},
\end{equation}

and the penalty function becomes

\begin{equation} \label{penalty_ancillas_ham}
P(x,y) = \sum_{i=1}^N \alpha_i \Bigg( \sum_{j=1}^N A_{ij} x_j - b_i - \sum_{\mu=0}^{\mu_i^m - 1} 2^{\mu} y_{i \mu} \Bigg)^2.
\end{equation}
The full classical Hamiltonian to be minimized is then

\begin{equation} \label{full_ham}
H(x,y) = C(x) + P(x,y).
\end{equation}
Since $H(x)$ only contains terms constant, linear, and quadratic in the binary variables $\{ x_i \}$ and $\{ y_{i \mu} \}$, it may be suitably programmed into the D-Wave as a BQM.

\section{Minor embedding of IEEE test power systems \label{embed}}

While reformulation of an optimization problem as a BQM is a necessary condition in order to be able to use AQA for its solution, it is not sufficient. One must also be able to embed the optimization problem into the hardware graph of the quantum processor, where nodes of the hardware graph represent physical qubits and edges represent physical qubit-qubit couplings. The 2000Q D-Wave quantum processing unit (QPU) is constructed from a 2,048 qubit (16,4)-chimera hardware graph topology consisting of a $16 \times 16$ array of 8-qubit chimera unit cells, each of which organized as a $K_{4,4}$ complete bipartite graph \cite{junger2019performance}. Meanwhile, a standard set of IEEE test power systems consists of the 9, 14, 24, 30, 39, 57, 118, and 300 bus test systems shown in the first column in Table \ref{scaling}. Column two shows the number of buses (nodes) in each test system while column three shows the corresponding number of branches (edges) in the test system. The number of ancilla bits required to represent Eq. \eqref{constraint} is shown in column four. The number of nontrivial pairwise interactions induced by the quadratic penalty function, Eq. \eqref{penalty_ancillas_ham}, is shown in column five. And finally, column six shows the minimum number of physical qubits found by the D-Wave Ocean API \textit{minorminer} tool required to embed the full problem Hamiltonian, Eq. \eqref{full_ham}, into the D-Wave 2000Q hardware graph by calling the \textit{find\_embedding} routine 10 times. The blank entry in column six denotes an instance where the embedding heuristic introduced by Cai et al. was unable to find a suitable embedding \cite{cai2014practical}. For this formulation of MDS and for the particular connectivity of the Chimera working graph, the number of physical qubits required to embed a given test system scales roughly linearly with the number of interactions induced by the penalty function. From this vantage point, it is clear why the IEEE 300 bus test system cannot be embedded in the D-Wave hardware graph, since its 3,478 interactions exceed the D-Wave's 2,048 qubits. This fact is confirmed by considering the instance of minor embedding a complete graph ($K_n$) in the (16, 4)-Chimera graph. It can be shown that the largest complete graph that can be embedded on the (16, 4)-Chimera graph is ($K_{\tilde{n}}$) where $\tilde{n} = 1+4\min(16,16)= 65$ \cite{junger2019performance}. A complete graph ($K_n$) has $n(n-1)/2$ edges, or interactions. For $\tilde{n}=65$, $\tilde{n}(\tilde{n}-1)/2=2,080$. Therefore, any interaction graph with a number of edges $>2,080$ cannot be embedded on the (16, 4)-Chimera architecture, and the 300 bus test system is ruled out.

\begin{table}
\caption{Scaling of resources for MDS on IEEE test power systems. \label{scaling}}
\begin{tabular}{ | c || c | c | c | c | c |}
\hline
System & Buses & Branches & Ancillas & Interactions & Qubits  \\
\hline\hline
IEEE 9 & 9 & 9 & 9 & 57 & 49 \\
\hline
IEEE 14 & 14 & 20 & 21 & 150 & 146 \\
\hline
IEEE 24 & 24 & 34 & 39 & 278 &  287 \\
\hline
IEEE 30 & 30 & 41 & 42 & 325 & 349 \\
\hline
IEEE 39 & 39 & 46 & 49 & 337 & 338 \\
\hline
IEEE 57 & 57 & 78 & 85 & 607 & 704 \\
\hline
IEEE 118 & 118 & 179 & 188 & 1,585 & 1,564 \\
\hline
IEEE 300 & 300 & 409 & 417 & 3,478 & - \\
\hline
\end{tabular}
\end{table}

\section{Assessment of solution quality \label{Sol_qual}}

Generally speaking, an optimization heuristic holds value if it has the prospect to reliably and quickly find optimal or near-optimal solutions to arbitrary, unstructured problem instances. With this perspective, the industry-standard CPLEX Optimizer can be regarded as an important benchmarking tool to make quantitative the terms ``reliably'' and ``quickly''. In order to further characterize the viability of current and near-term quantum optimization heuristics for power system design, we compare both the best solution found and time to best solution from the D-Wave 2000Q processor with CPLEX Optimizer results for the ILP in Eqs. \eqref{cost} and \eqref{constraint} obtained in recent work by K. Sou \cite{sou2016branch}. We do not regard any of the numerically timed quantities as immutable fact since the results of numerical experiments generally depend not only upon the algorithms at play but also the particular hardware on which they are run and their implementation. It is nevertheless illustrative to compare the scaling of 2000Q performance against a relatively standard optimization package running on relatively standard hardware.

We report two times-to-best metrics for the quantum annealer. A full quantum computation to find an optimal solution takes time
\begin{equation}
T = T_P + k(\tau+T_R),
\end{equation}
%
\begin{table}
\caption{Comparison of solutions and computation times (in seconds) for MDS on IEEE test power systems for SA, CPLEX, and AQA. \label{solutions}}
\centering
\begin{tabular}{ |c||c|c|c|c|c|}
\hline
System & $\gamma_{SA}$ & $\gamma_{AQA}$ & $T_{CPLX} (s) $ \cite{sou2016branch} & $T_{A} (s)$ & $T (s)$ \\
\hline\hline
IEEE 9 & 3 & 3 & 0.0016 & 0.000070 & 0.0094 \\
\hline
IEEE 14 & 4 & 4 & 0.0066 & 0.000004 & 0.0086 \\
\hline
IEEE 24 & 7 & 7 & 0.0083 & 0.039671 & 0.2158 \\
\hline
IEEE 30 & 10 & 10 & 0.0063 & 0.000342 & 0.0130 \\
\hline
IEEE 39 & 13 & 14 & 0.0065 & 0.281507 & 0.4500 \\
\hline
IEEE 57 & 17 & 20 & 0.0140 & 0.078621 & 0.0910 \\
\hline
IEEE 118 & 32 & 49 & 0.0100 & 2.985983 & 4.4380 \\
\hline
\end{tabular}
\end{table}
where $T_P$ is the time required to program the problem onto the QPU and initialize the control sequence, $\tau$ (introduced in Sec. \ref{quantop}) is the time taken for one annealing schedule to complete, $T_R$ is the time it takes to read a measurement at the end of one annealing schedule, and $k$ is the number of times the anneal-read cycle is repeated in order to ideally obtain adequate statistics on the probabilities $\{|a_i|^2\}$. We call $T_A \equiv k \tau$ the ``annealing time'' and $T$ the ``QPU access time''. It is conventional to consider $T_A$ as the quantum equivalent to classical CPU time since $T_P$ has its classical equivalent in the time it takes to compile classical code and $T_R$ is fixed overhead for any quantum computation and therefore does not give any useful information about the scaling of the AQA algorithm proper \cite{king2019quantum}. However, we do report on $T$ as well for completeness. Column one of Table \ref{solutions} again shows the IEEE test system analyzed. Column two displays best-found domination numbers obtained by SA $(\gamma_{SA})$, which corroborate those found in Ref. \cite{sou2016branch} using CPLEX. Column three shows the best-found domination number obtained by the 2000Q processor $(\gamma_{AQA})$. Column four shows the time to best solution for the CPLEX Optimizer $(T_{CPLX})$ running on a Mac with a 2.5 GHz CPU and 8GB of RAM. Column five shows the annealing time to best solution $(T_A)$ while column six shows the corresponding QPU access time $(T)$.

For our AQA calculations we set the penalty strength $\alpha_i = 2 \: \: \forall \: i$, which corresponds to a reasonably hard enforcement of Eq. \eqref{constraint} in the penalty function-- Eq. \eqref{penalty_ancillas_ham}-- and the chain strength $J_{chain} = 1.5 |J|_M$, which gauges how strongly different physical qubits representing a single logical qubit interact. $|J|_M$ is the maximum coupling strength encountered in the Ising formulation of each problem instance. Note that $|J|_M = 8$ for all test systems except IEEE 9 for which $|J|_M=2$ and IEEE 118 and 300 for which $|J|_M=32$. $J_{chain}$ was chosen such that the fraction of logical of chains broken upon readout of best-found solutions was bounded from above by $0.001$. $T_A$ was calculated in the following manner. Grids of $\tau$ and $k$ values were created: $\tau \in [1 \mu s, 1728 \mu s]$ and $k \in [1, 1728]$, so that the maximum annealing time considered was less than $3 s$-- the maximum annealing time allowed by the 2000Q. For each parameter, 20 grid points were chosen, evenly spaced on a base-12 logarithmic scale, and rounded to the nearest integer. At each point in the $\{\tau\} \times \{k\}$ grid, the lowest energy solution to Eq. \eqref{full_ham} found by the corresponding annealing schedule was obtained, $x^*(\tau,k)$ and only solutions that satisfy Eq. \eqref{constraint} were kept. For a given IEEE test system then, $\gamma_{AQA} = \min_{(\tau,k)} \sum_i x_i^*(\tau,k)$ and $T_A = k^*\times \tau^*$ where $(\tau^*,k^*)= \argmin \gamma_{AQA}$.

As can be seen in columns one and two of Table \ref{solutions}, the 2000Q processor is able to find the same domination number as obtained by SA and CPLEX in the four smallest problem instances of the seven for which a suitable minor embedding was found. In three of four of these instances (IEEE 9, 14, and 30), the $T_A$ is at least an order of magnitude less than $T_{CPLX}$. Interestingly, these three test systems are all planar graphs. There are two non-planar problem graphs for which a minor embedding was found (IEEE 24, and 57). While $\gamma_{SA}$ was found by the 2000Q for IEEE 24, $T_A$ was roughly five times larger than $T_{CPLX}$ on that problem instance. For IEEE 57, the correct domination number is missed by three additional PMUs and $T_A \sim 6 \: T_{CPLX}$. Finally, while the 2000Q misses $\gamma_{SA}$ by only one additional PMU for IEEE 39, the solution quality deteriorates rapidly for the larger problem instance of IEEE 118. The failure at this problem instance points to an important improvement to be made in future AQA hardware. IEEE 118 is the only embeddable problem instance in which $|J|_M=32$ and $|h|_M=56$. We conjecture that when autoscaled to fit within the machine-specific analog parameter ranges $h \in [-2,2]$ and $J \in [-1,1]$ (or $J \in [-2,1]$ for the VFYC solver), the resulting hardware-level energy gaps between different solutions become small compared to system temperature $k_B T_{sys}$. And while decreasing $J_{chain}$ to $1.0|J|_M$ for IEEE 118 resulted in an improvement in time to solution: $T_A \sim 0.29s$ and $T\sim 0.44s$, the best found solution was similarly poor $\gamma_{AQA}=49$. Hence, extending parameter ranges and ensuring that energy gaps can be made large enough with respect to system temperature to avoid thermal-noise for large and highly-connected problem instances should be a main focus of next-generation hardware design.

\section{Conclusion}

The results presented in Sec. \ref{Sol_qual} corroborate that AQA holds the potential to outpace well-developed classical optimization methods on combinatorial optimization problems. Overcoming the current limitations in hardware connectivity and thermal noise for large and highly-connected graphs should allow quantum optimization to begin to address increasingly challenging problems, and therefore become a useful tool in the design of future power systems.



\section*{Acknowledgment}

This work was authored in part by the National Renewable Energy Laboratory (NREL), operated by Alliance for Sustainable Energy, LLC, for the U.S. Department of Energy (DOE) under Contract No. DE-AC36-08GO28308. This work was supported by the Laboratory Directed Research and Development (LDRD) Program at NREL. The views expressed in the article do not necessarily represent the views of the DOE or the U.S. Government. The U.S. Government retains and the publisher, by accepting the article for publication, acknowledges that the U.S. Government retains a nonexclusive, paid-up, irrevocable, worldwide license to publish or reproduce the published form of this work, or allow others to do so, for U.S. Government purposes. This research used Ising, Los Alamos National Laboratory's D-Wave quantum annealer.  Ising is supported by NNSA's Advanced Simulation and Computing program. The authors would like to thank Scott Pakin and Denny Dahl. This material is based in-part upon work supported by the National Science Foundation under Grant No. PHY-1653820.



\bibliographystyle{IEEEtran}
%



\end{document}